\documentstyle[12pt]{article}
\newlength{\extraspace}
\newlength{\extraspaces}
\newcommand{\be}{\begin{equation}
\addtolength{\abovedisplayskip}{\extraspaces}
\addtolength{\belowdisplayskip}{\extraspaces}
\addtolength{\abovedisplayshortskip}{\extraspace}
\addtolength{\belowdisplayshortskip}{\extraspace}}
\newcommand{\ee}{\end{equation}}
\newcommand{\ba}{\begin{eqnarray}
\addtolength{\abovedisplayskip}{\extraspaces}
\addtolength{\belowdisplayskip}{\extraspaces}
\addtolength{\abovedisplayshortskip}{\extraspace}
\addtolength{\belowdisplayshortskip}{\extraspace}}
\newcommand{\ea}{\end{eqnarray}}
\newcommand{\nonu}{\nonumber \\[.5mm]}
\newcommand{\A}{&\!\!\!}
\begin{document}
\thispagestyle{empty}
\begin{flushright}
SIT-LP-08/05 \\
%{\tt arXiv:yymm.nnnn[hep-th]} \\
May, 2008
\end{flushright}
\vspace{7mm}
%
%\vspace*{7mm}
%
\begin{center}
{\large{\bf On Wess-Zumino gauge}} \\[20mm]
{\sc Kazunari Shima}
\footnote{
\tt e-mail: shima@sit.ac.jp} \ 
and \ 
{\sc Motomu Tsuda}
\footnote{
\tt e-mail: tsuda@sit.ac.jp} 
\\[5mm]
{\it Laboratory of Physics, 
Saitama Institute of Technology \\
Fukaya, Saitama 369-0293, Japan} \\[20mm]
\begin{abstract}
In the relation between the linear (L) supersymmetry (SUSY) representation 
and the nonlinear (NL) SUSY representation we discuss the role of the Wess-Zumino gauge. 
We show in two dimensional spacetime that a spontaneously broken LSUSY theory 
with mass and Yukawa interaction terms for a minimal off-shell vector supermultiplet 
is obtained from a general superfield without imposing any special gauge conditions 
in $N = 2$ NL/L SUSY relation. 
\\[5mm]
\noindent
PACS: 11.30.Pb, 12.60.Jv, 12.60.Rc, 12.10.-g \\[2mm]
\noindent
Keywords: supersymmetry, superfield, Nambu-Goldstone fermion, 
nonlinear/linear SUSY relation, composite unified theory 
\end{abstract}
\end{center}

\newpage

\noindent
Superfields on superspace provide a method to formulate SUSY field theories 
systematically \cite{WB,PW}. 
The Wess-Zumino (WZ) gauge is used conveniently to eliminate 
redundant (gauge) degrees of freedom (d.o.f.) from general superfields for vector supermultiplets. 
Linear (L) SUSY actions for minimal off-shell vector supermultiplets (in interacting theories) 
are constructed by using superfields in the WZ gauge. 

On the other hand, various LSUSY theories are related (equivalent) to 
a NL SUSY model \cite{VA} through the linearization of NLSUSY 
both for the free case \cite{IK}-\cite{STT2} 
and for the interacting (SUSY-QED) case \cite{ST1,ST2}, 
which are essential to study the low energy physics and cosmology 
of NLSUSY general relativity (GR) \cite{KS} in the SGM scenario \cite{ST3}. 
In the NL/L SUSY relation, all component fields for the vector supermultiplets 
including the redundant ones of the general superfields are consistently expressed 
as composites of Nambu-Goldstone (NG) fermion \cite{STT1,ST4} (called {\it superon} in the SGM scenario), 
whose expressions are called {\it SUSY invariant relations}. 
This means the WZ gauge condition used in LSUSY theories gives apparently 
the artificial (gauge dependent) restrictions on the composite states of NG fermions, 
which looks unnatural so far provided the composites are regarded as some eigen state of spacetime symmetry. 
Therefore, it is worthwhile studying the general properties of the superfield formulation 
which does not rely on the special gauge condition in viewpoints of NL/L SUSY relation. 

In this letter, we focus on the interacting SUSY theory for the $N = 2$ vector supermultiplet 
in two dimensional spacetime ($d = 2$) for simplicity.  
Note that $N = 1$ vector supermultiplet is unphysical \cite{STL} in the SGM scenario. 
We show that by using the $N = 2$ NL/L SUSY relation constructed systematically 
in the previous paper \cite{ST4} 
the interacting SUSY theory for the minimal off-shell vector supermultiplet and 
for the  general vector supermultiplet as well 
accompanying the spontaneous SUSY breaking are obtained from the general superfield 
without imposing any special gauge conditions. 

The $N = 2$ general superfield on superspace coordinates $(x^a, \theta^i)$ 
for the $N = 2$ LSUSY vector supermultiplet (in $d = 2$) is given by \cite{DVF,ST5} 
\ba
{\cal V}(x, \theta^i) \A = \A C(x) + \bar\theta^i \Lambda^i(x) 
+ {1 \over 2} \bar\theta^i \theta^j M^{ij}(x) 
- {1 \over 2} \bar\theta^i \theta^i M^{jj}(x) 
+ {1 \over 4} \epsilon^{ij} \bar\theta^i \gamma_5 \theta^j \phi(x) 
\nonu
\A \A 
- {i \over 4} \epsilon^{ij} \bar\theta^i \gamma_a \theta^j v^a(x) 
- {1 \over 2} \bar\theta^i \theta^i \bar\theta^j \lambda^j(x) 
- {1 \over 8} \bar\theta^i \theta^i \bar\theta^j \theta^j D(x), 
\label{VSF}
\ea
where $M^{ij} = M^{(ij)}$ $\left(= {1 \over 2}(M^{ij} + M^{ji}) \right)$ 
and $M^{ii} = \delta^{ij} M^{ij}$. 
LSUSY transformations for the (general) component fields $(C, \Lambda^i, M^{ij}, \cdots)$ 
with constant (Majorana) spinor parameters $\zeta^i$ 
are deduced from the following supertranslation of the superfield in the superspace $(x^a, \theta^i)$, 
\be
\delta_\zeta {\cal V}(x, \theta^i) = \bar\zeta^i Q^i {\cal V}(x, \theta^i) 
\label{VSFtransfn}
\ee
with supercharges $Q^i = {\partial \over \partial\bar\theta^i} + i \!\!\not\!\partial \theta^i$. 
Gauge transformations for $(C, \Lambda^i, M^{ij}, \cdots)$ are also defined as follows 
by means of \cite{ST5} 
\be
\delta_g {\cal V}(x, \theta^i) = \Phi^1(x, \theta^i) + \Phi^2(x, \theta^i), 
\label{gauge}
\ee
where $\Phi^i$ ($i = 1, 2$) are two scalar superfield parameters for generalized gauge parameters 
(for a $N = 2$ matter scalar supermultiplet (for example, see \cite{UZ})). 

As in the case for $d = 4$, $N = 1$ SUSY \cite{WB}, 
gauge invariant quantities for $N = 2$ denoted tentatively as $(A_0, \phi_0, F_{0ab}, \lambda_0, D_0)$ 
can be easily constructed by the explicit component form 
of the gauge transformation (\ref{gauge}) as follows; 
\be
(A_0, \phi_0, F_{0ab}, \lambda_0^i, D_0) 
\equiv (M^{ii}, \phi, F_{ab}, \lambda^i + i \!\!\not\!\partial \Lambda^i, D + \Box C), 
\label{minimal}
\ee
where $F_{0ab} = \partial_a v_{0b} - \partial_b v_{0a}$, $F_{ab} = \partial_a v_b - \partial_b v_a$ 
and $v_{0a} = v_a$ transforms as an Abelian gauge field.  
The familiar LSUSY transformations are calculated directly from Eq.(\ref{VSFtransfn}). 
And for the minimal off-shell vector supermultiplet 
$(A_0, \phi_0, v_{0a}, \lambda_0^i, D_0)$ \cite{ST1}, we obtain 
\ba
\A \A 
\delta_\zeta A_0 = \bar\zeta^i \lambda_0^i, 
\nonu
\A \A 
\delta_\zeta \phi_0 = - \epsilon^{ij} \bar\zeta^i \gamma_5 \lambda_0^j, 
\nonu
\A \A 
\delta_\zeta v_{0a} = - i \epsilon^{ij} \bar\zeta^i \gamma_a \lambda_0^j + \partial_a W(\Lambda^i), 
\nonu
\A \A 
\delta_\zeta \lambda_0^i = (D_0 - i \!\!\not\!\partial A_0) \zeta^i 
- i \epsilon^{ij} \gamma_5 \!\!\not\!\partial \phi_0 \zeta^j 
+ {1 \over 2} \epsilon^{ab} \epsilon^{ij} F_{0ab} \gamma_5 \zeta^j, 
\nonu
\A \A 
\delta_\zeta D_0 = - i \bar\zeta^i \!\!\not\!\partial \lambda_0^i, 
\label{LSUSY0}
\ea
where remarkably the redundant components in Eq.(\ref{VSF}), $(C, \Lambda^i, M^{12}, M^{11} - M^{22})$, 
do not appear apparently in the minimal off-shell vector multiplet 
$\delta_\zeta (A_0, \phi_0, v_{0a}, \lambda_0^i, D_0)$ disregarding the gauge choice 
except for the $U(1)$ gauge parameter $W(\Lambda^i) = - 2 \epsilon^{ij} \bar\zeta^i$ $\Lambda^j$ 
composed of the fermionic d.o.f.. 
That is, the gauge invariant sector and the redundant sector are mixed by only the gauge transformations 
with the fermionic parameters. 
Therefore the redundant sector plays important roles for constructing SUSY QED.  

The LSUSY transformations (\ref{LSUSY0}) satisfy the closed commutator algebra, 
\be
[ \delta_{\zeta_1}, \delta_{\zeta_2} ] = \delta_P(\Xi^a), 
\label{comm}
\ee
where $\delta_P(\Xi^a)$ means a translation with a parameter $\Xi^a = 2 i \bar\zeta_1^i \gamma^a \zeta_2^i$. 
Interestingly, in contrast to the WZ gauge, the commutator algebra for $v_{0a}$ does not contain 
the $U(1)$ gauge transformation term due to 
\be
\delta_{\zeta_1} W_{\zeta_2}(\Lambda^{i}) - \delta_{\zeta_2} W_{\zeta_1}(\Lambda^{i}) 
= - 2 (\epsilon^{ij} \bar\zeta_1^i \zeta_2^j A_0 
+ \bar\zeta_1^i \gamma_5 \zeta_2^i \phi_0
- i \bar\zeta_1^i \gamma^a \zeta_2^i v_{0a}) 
\ee
and the consequent cancellations. 

Now we see the above general arguments in the viewpoints of NL/L SUSY relation. 
Let us introduce the SUSY invariant relations between $(C, \Lambda^i, M^{ij}, \cdots)$ 
of $N = 2$ LSUSY and (Majorana) NG fermions $\psi^i$ of $N = 2$ NLSUSY \cite{ST4} in $d = 2$, 
in which each component field is expanded in terms of $\psi^i$ 
with respect to a constant $\kappa$ whose dimension is $({\rm mass})^{-1}$ 
($\kappa^{-2} \sim {\Lambda \over G}$ in NLSUSY GR \cite{KS}) as 
\be
(C, \Lambda^i, M^{ij}, \cdots) \sim \xi \kappa^{n-1} (\psi^i)^n \vert w \vert \ (n = 4, 3, \cdots, 0), 
\label{SUSYinv0}
\ee
where $\xi$ is an arbitrary dimensionless constant, 
$(\psi^i)^4 = \bar\psi^i \psi^i \bar\psi^j \psi^j$, $(\psi^i)^3 = \psi^i \bar\psi^j \psi^j$ 
and $(\psi^i)^2 = \bar\psi^i \psi^j, \epsilon^{ij} \bar\psi^i \gamma_5 \psi^j, 
\epsilon^{ij} \bar\psi^i \gamma^a \psi^j$, 
which are very promissing features for SGM scenario \cite{ST3}. 
In Eq.(\ref{SUSYinv0}), $\vert w \vert$ is the determinant introduced in \cite{VA} 
describing the dynamics of NG fermions of NLSUSY, 
i.e. for the $d = 2$, $N = 2$ ($N > 2$, as well) NLSUSY case, 
\be
\vert w \vert = \det(w^a{}_b) = \det(\delta^a_b + t^a{}_b), 
\ \ \ t^a{}_b = - i \kappa^2 \bar\psi^i \gamma^a \partial_b \psi^i, 
\ee
expanded in terms of $t^a{}_b$ or $\psi^i$ as 
\ba
\vert w \vert \A = \A 1 + t^a{}_a + {1 \over 2!}(t^a{}_a t^b{}_b - t^a{}_b t^b{}_a) 
\nonu
\A = \A 1 - i \kappa^2 \bar\psi^i \!\!\not\!\partial \psi^i 
- {1 \over 2} \kappa^4 
(\bar\psi^i \!\!\not\!\partial \psi^i \bar\psi^j \!\!\not\!\partial \psi^j 
- \bar\psi^i \gamma^a \partial_b \psi^i \bar\psi^j \gamma^b \partial_a \psi^j) 
\nonu
\A = \A 1 - i \kappa^2 \bar\psi^i \!\!\not\!\partial \psi^i 
- {1 \over 2} \kappa^4 \epsilon^{ab} 
(\bar\psi^i \psi^j \partial_a \bar\psi^i \gamma_5 \partial_b \psi^j 
+ \bar\psi^i \gamma_5 \psi^j \partial_a \bar\psi^i \partial_b \psi^j). 
\ea
Note that the NG fermion fields $\psi^i$ transform nonlinearly as 
\be
\delta_\zeta \psi^i = {1 \over \kappa} \zeta^i 
- i \kappa \bar\zeta^j \gamma^a \psi^j \partial_a \psi^i, 
\label{NLSUSY}
\ee
which also satisfy the commutator algebra (\ref{comm}). 

Explicit forms of the SUSY invariant relations in all orders of $\psi^i$ of Eq.(\ref{SUSYinv0}) are given as follows; 
\ba
C \A = \A - {1 \over 8} \xi \kappa^3 \bar\psi^i \psi^i \bar\psi^j \psi^j, 
\nonu
\Lambda^i \A = \A - {1 \over 2} \xi \kappa^2 
\psi^i \bar\psi^j \psi^j (1 - i \kappa^2 \bar\psi^k \!\!\not\!\partial \psi^k), 
\nonu
M^{ij} \A = \A {1 \over 2} \xi \kappa \bar\psi^i \psi^j 
\left( 1 - i \kappa^2 \bar\psi^k \!\!\not\!\partial \psi^k 
- {1 \over 2} \kappa^4 \epsilon^{ab} \bar\psi^k \psi^l 
\partial_a \bar\psi^k \gamma_5 \partial_b \psi^l \right), 
\nonu
\phi \A = \A - {1 \over 2} \xi \kappa \epsilon^{ij} \bar\psi^i \gamma_5 \psi^j 
\left( 1 - i \kappa^2 \bar\psi^k \!\!\not\!\partial \psi^k 
- {1 \over 2} \kappa^4 \epsilon^{ab} \bar\psi^k \gamma_5 \psi^l 
\partial_a \bar\psi^k \partial_b \psi^l \right), 
\nonu
v^a \A = \A - {i \over 2} \xi \kappa \epsilon^{ij} \bar\psi^i \gamma^a \psi^j 
(1 - i \kappa^2 \bar\psi^k \!\!\not\!\partial \psi^k), 
\nonu
\lambda^i \A = \A \xi \psi^i \vert w \vert, 
\nonu
D \A = \A {\xi \over \kappa} \vert w \vert. 
\label{SUSYinv}
\ea

The SUSY invariant relations (\ref{SUSYinv}) are systematically obtained \cite{ST4} 
by considering the general superfield, 
\be
\tilde{\cal V}(x, \theta^i) = {\cal V}(x', \theta'^i), 
\label{VSF'}
\ee
on the specific supertranslations $(x'^a, \theta'^i)$ \cite{IK,UZ} depending on $\psi^i$, 
\ba
\A \A 
x'^a = x^a + i \kappa \bar\theta^i \gamma^a \psi^i, 
\nonu
\A \A 
\theta'^i = \theta^i - \kappa \psi^i. 
%\label{specific}
\ea
Expanding the superfield (\ref{VSF'}) in the power series of $\theta^i$, i.e. 
\ba
\tilde{\cal V}(x, \theta^i) \A = \A \tilde C(x) + \bar\theta^i \tilde\Lambda^i(x) 
+ {1 \over 2} \bar\theta^i \theta^j \tilde M^{ij}(x) 
- {1 \over 2} \bar\theta^i \theta^i \tilde M^{jj}(x) 
+ {1 \over 4} \epsilon^{ij} \bar\theta^i \gamma_5 \theta^j \tilde\phi(x) 
\nonu
\A \A 
- {i \over 4} \epsilon^{ij} \bar\theta^i \gamma_a \theta^j \tilde v^a(x) 
- {1 \over 2} \bar\theta^i \theta^i \bar\theta^j \tilde\lambda^j(x) 
- {1 \over 8} \bar\theta^i \theta^i \bar\theta^j \theta^j \tilde D(x), 
\ea
and ${\cal V}(x', \theta'^i)$ as well and imposing SUSY invariant constraints, 
\be
\tilde C = \tilde\Lambda^i = \tilde M^{ij} = \tilde\phi = \tilde v^a = \tilde\lambda^i = 0, 
\ \ \ \tilde D = {\xi \over \kappa}, 
\label{const}
\ee
give the SUSY invariant relations (\ref{SUSYinv}). 
Note that the SUSY invariant constraints (\ref{const}) do not restrict the NG fermion-composite structure 
of LSUSY multiplet fields. 

The SUSY invariant relations for the component fields 
($A_0$, $\phi_0$, $v_{0a}$, $\lambda_0$, $D_0$) constituting the minimal off-shell vector supermultiplet 
are constructed from Eq.(\ref{SUSYinv}). 
And they coincide with the results obtained heuristicly in Ref.\cite{ST1}, 
which induce the SUSY transformations (\ref{LSUSY0}) and the commutator algebra (\ref{comm}) 
under the NLSUSY transformations (\ref{NLSUSY}). 

Now the problem is actions constructed from the general superfield (\ref{VSF}) in NL/L SUSY relation. 
The $N = 2$ LSUSY free (kinetic) action with a Fayet-Iliopulos (FI) term is written as 
\ba
S_{{\cal V}{\rm free}} \A = \A \int d^2 x \left\{ 
\int d^2 \theta^i \ {1 \over 32} (\overline{D^i {\cal W}^{jk}} D^i {\cal W}^{jk} 
+ \overline{D^i {\cal W}_5^{jk}} D^i {\cal W}_5^{jk}) 
+ \int d^4 \theta^i \ {\xi \over {2 \kappa}} {\cal V} 
\right\}_{\theta^i = 0} 
\nonu
\A = \A \int d^2 x \left\{ - {1 \over 4} (F_{0ab})^2 
+ {i \over 2} \bar\lambda_0^i \!\!\not\!\partial \lambda_0^i 
+ {1 \over 2} (\partial_a A_0)^2 + {1 \over 2} (\partial_a \phi_0)^2 + {1 \over 2} D_0^2 
\right. 
\nonu
\A \A 
\hspace*{1.4cm} \left. - {\xi \over \kappa} (D_0 - \Box C) \right\} 
\nonu
\A = \A S^0_{{\cal V}{\rm free}} + \int d^2 x \ {\xi \over \kappa} \Box C, 
\label{free}
\ea
where ${\cal W}^{ij}$ and ${\cal W}_5^{ij}$ are scalar and pseudo scalar superfields defined by 
\be
{\cal W}^{ij} = \bar D^i D^j {\cal V}, \ \ \ {\cal W}_5^{ij} = \bar D^i \gamma_5 D^j {\cal V} 
\ee
with $D^i = {\partial \over \partial\bar\theta^i} - i \!\!\not\!\partial \theta^i$. 
Obviously, the free action (\ref{free}) is expressed by only the fields 
($A_0$, $\phi_0$, $v_{0a}$, $\lambda_0$, $D_0$) except the surface term without choosing the specific gauge. 
By changing the integration variables in Eq.(\ref{free}) from $(x, \theta^i)$ to $(x', \theta'^i)$ 
under the SUSY invariant constraints (\ref{const}) \cite{ST4} 
or by directly substituting the SUSY invariant relations (\ref{SUSYinv}) into Eq.(\ref{free}) \cite{ST1}, 
$S_{{\cal V}{\rm free}}$ exactly reduces to the $N = 2$ NLSUSY action, i.e. 
\ba
S_{{\cal V}{\rm free}}(\psi^i) = S_{N = 2{\rm NLSUSY}}, 
\label{free-NLSUSY}
\ea
when $\xi^2 = 1$, where $S_{N = 2{\rm NLSUSY}}$ is 
\be
S_{N = 2{\rm NLSUSY}} = - {1 \over {2 \kappa^2}} \int d^2 x \ \vert w \vert, 
\label{NLSUSYaction}
\ee
which is invariant (becomes a surface term) under the NLSUSY transformations (\ref{NLSUSY}). 

Next we focus on mass and Yukawa interaction terms constructed from Eq.(\ref{VSF}). 
We define those terms by the quadratic and the cubic terms of ${\cal W}^{ij}$ and ${\cal W}_5^{ij}$ 
as follows; 
\ba
S_{{\cal V}m} \A = \A {1 \over 16} \int d^2 x \ m \left[ 
\int d^2 \theta^i \ 3 \{ ({\cal W}^{jk})^2 + ({\cal W}_5^{jk})^2 \} 
\right. 
\nonu
\A \A 
\hspace{2.2cm} \left. 
+ \int d \bar\theta^i d \theta^j 
\ 2 ({\cal W}^{ik} {\cal W}^{jk} + {\cal W}_5^{ik} {\cal W}_5^{jk}) 
\right]_{\theta^i = 0}, 
\label{mass}
\\
S_{{\cal V}f} \A = \A {1 \over 8} \int d^2 x \ f \left[ 
\int d^2 \theta^i \ {\cal W}^{jk} 
({\cal W}^{jl} {\cal W}^{kl} + {\cal W}_5^{jl} {\cal W}_5^{kl}) 
\right. 
\nonu
\A \A 
+ \int d \bar\theta^i d \theta^j \ 2 \{ {\cal W}^{ij} 
({\cal W}^{kl} {\cal W}^{kl} + {\cal W}_5^{kl} {\cal W}_5^{kl}) 
\nonu
\A \A 
\hspace{2.2cm} 
+ {\cal W}^{ik} ({\cal W}^{jl} {\cal W}^{kl} + {\cal W}_5^{jl} {\cal W}_5^{kl}) \} 
\bigg]_{\theta^i = 0}, 
\label{Yukawa}
\ea
where $f$ is an arbitrary constant with the dimension (mass)$^1$. 
The actions (\ref{mass}) and (\ref{Yukawa}) contain the redundant fields in ${\cal V}(x, \theta^i)$ 
but these actions give the $N = 2$ LSUSY mass and the Yukawa interaction terms 
for the minimal off-shell vector supermultiplet (by imposing the WZ gauge condition) \cite{ST1}. 
In NL/L SUSY relation, it can be easily shown that the actions (\ref{mass}) and (\ref{Yukawa}) 
vanish, i.e. 
\be
S_{{\cal V}m}(\psi^i) = 0, \ \ \ S_{{\cal V}f}(\psi^i) = 0, 
\label{vanish}
\ee
by changing the integration variables in Eqs.(\ref{mass}) and (\ref{Yukawa}) 
from $(x, \theta^i)$ to $(x', \theta'^i)$ under the SUSY invariant constraints (\ref{const}). 

The results (\ref{vanish}) mean the followings. 
The explicit (general) component form of Eqs.(\ref{mass}) and (\ref{Yukawa}) become 
\ba
S_{{\cal V}m} \A = \A \int d^2 x \ m 
\left\{ - {1 \over 2} (\bar\lambda_0^i \lambda_0^i - 2 A_0 D_0 + \epsilon^{ab} \phi_0 F_{0ab}) 
- 2 (2 A_0 \Box C - \partial_a \bar\Lambda^i \gamma^a \!\!\not\!\partial \Lambda^i) 
\right\} 
\nonu
\A = \A S^0_{{\cal V}m} 
- 2 \int d^2 x \ m (2 A_0 \Box C - \partial_a \bar\Lambda^i \gamma^a \!\!\not\!\partial \Lambda^i), 
\label{mass-comp}
\\
%\A = \A 0 \ \ \ (\ {\rm in\ NLSUSY\ rep.}; \ S^0_{{\cal V}m} = 0\ )\ . 
S_{{\cal V}f} \A = \A \int d^2 x \ f 
[ ( A_0 \bar\lambda_0^i \lambda_0^i + \epsilon^{ij} \phi_0 \bar\lambda_0^i \gamma_5 \lambda_0^j 
- A_0^2 D_0 + \phi_0^2 D_0 + \epsilon^{ab} A_0 \phi_0 F_{0ab} )  
\nonu
\A \A 
\hspace{1.7cm}
+ \{ A_0 \bar\lambda^i \lambda^i - 2 M^{ij} \bar\lambda^i \lambda^j 
- 2 A_0^2 D + 4 (M^{ij})^2 D + \cdots \} ] 
\nonu
\A = \A S^0_{{\cal V}f} 
+ \int d^2 x \ f \{ A_0 \bar\lambda^i \lambda^i - 2 M^{ij} \bar\lambda^i \lambda^j 
- 2 A_0^2 D + 4 (M^{ij})^2 D + \cdots \}, 
%\nonu
%\A = \A 0 \ \ \ (\ {\rm in\ NLSUSY\ rep.}; \ S^0_{{\cal V}f} = 0\ )\ . 
\label{Yukawa-comp}
\ea
where $S^0_{{\cal V}m}$ and $S^0_{{\cal V}f}$ are the mass and the Yukawa interaction terms 
for the minimal off-shell vector supermultiplet, 
which are invariant under the LSUSY transformations (\ref{LSUSY0}), respectively. 
By directly substituting Eq.(\ref{SUSYinv}) into $S^0_{{\cal V}m}$ and $S^0_{{\cal V}f}$, 
the (nontrivial) vanishments of the actions, 
\be
S^0_{{\cal V}m}(\psi^i) = 0, \ \ \ S^0_{{\cal V}f}(\psi^i) = 0, 
\label{vanish0}
\ee
are shown by means of miraculous cancellations among four NG fermion self- interaction terms \cite{ST1}. 
Therefore, it is understood that the redundant terms 
in the actions (\ref{mass-comp}) and (\ref{Yukawa-comp}) 
also vanish by means of the cancellations among terms, which can be confirmed easily. 

The above arguments show that, under the SUSY invariant relations (\ref{SUSYinv}), 
the $N = 2$ LSUSY action with the mass and the Yukawa interaction terms 
for the minimal off-shell vector supermultiplet ($A_0$, $\phi_0$, $v_{0a}$, $\lambda_0$, $D_0$) 
is obtained from the NLSUSY action (\ref{NLSUSYaction}) 
without imposing any special gauge conditions for the general superfield (\ref{VSF}); 
namely, the (nontrivial) vanishment of the redundant (subsidiary) structure of the action 
in NL/L SUSY relation in addition to Eqs.(\ref{free-NLSUSY}), (\ref{vanish}) and (\ref{vanish0}) 
leads to 
\ba
S_{N = 2{\rm NLSUSY}} \A = \A S_{{\cal V}{\rm free}} + S_{{\cal V}m} + S_{{\cal V}f} 
\nonu
\A = \A S^0_{{\cal V}{\rm free}} + S^0_{{\cal V}m} + S^0_{{\cal V}f} + [\ {\rm surface\ term}\ ]. 
\label{NL/LSUSY}
\ea

To summarize, based on the general superfield (\ref{VSF}), 
we have discussed the relation between the $N = 2$ NLSUSY model and the $N = 2$ LSUSY theory 
for the vector supermultiplet with the mass and the Yukawa interaction terms in $d = 2$. 
The SUSY invariant relations and the SUSY invariant constraints in NL/L SUSY relation 
are expressed on the most general supermultiplet of the vector superfield. 
The SUSY invariant relations (\ref{SUSYinv}), which are systematically obtained \cite{ST4} 
from the SUSY invariant constraints (\ref{const}), are compatible with 
the gauge invariances of the quantities (\ref{minimal}) for the minimal off-shell vector supermultiplet \cite{ST1}. 
As for the actions constructed from Eq.(\ref{VSF}), in particular, 
the mass and the Yukawa interaction terms (\ref{mass}) and (\ref{Yukawa}), 
we have shown that the redundant (subsidiary) structure of the actions vanishes in NL/L SUSY relation 
in addition to Eqs.(\ref{free-NLSUSY}), (\ref{vanish}) and (\ref{vanish0}). 
This means Eq.(\ref{NL/LSUSY}), 
i.e. the $N = 2$ LSUSY action with the mass and the Yukawa interaction terms for the minimal off-shell vector supermultiplet 
emerges from the NLSUSY action (\ref{NLSUSYaction}) 
without imposing any special gauge conditions for the general superfield (\ref{VSF}). 

The same structure as the $N = 2$ NL/L SUSY relation is anticipated for higher $N$ SUSY 
and in $d = 4$ as well. 
The similar arguments in the coupling of matter multiplet (SUSY QED) and in the SUSY QCD  
are also interesting problems in NL/L SUSY relation and are under the study.

\newpage

%%%%%%%  References  %%%%%%%%%%%%%%%%%%%%%%%%%%%%%%%%%%%%%%%
%
\newcommand{\NP}[1]{{\it Nucl.\ Phys.\ }{\bf #1}}
\newcommand{\PL}[1]{{\it Phys.\ Lett.\ }{\bf #1}}
\newcommand{\CMP}[1]{{\it Commun.\ Math.\ Phys.\ }{\bf #1}}
\newcommand{\MPL}[1]{{\it Mod.\ Phys.\ Lett.\ }{\bf #1}}
\newcommand{\IJMP}[1]{{\it Int.\ J. Mod.\ Phys.\ }{\bf #1}}
\newcommand{\PR}[1]{{\it Phys.\ Rev.\ }{\bf #1}}
\newcommand{\PRL}[1]{{\it Phys.\ Rev.\ Lett.\ }{\bf #1}}
\newcommand{\PTP}[1]{{\it Prog.\ Theor.\ Phys.\ }{\bf #1}}
\newcommand{\PTPS}[1]{{\it Prog.\ Theor.\ Phys.\ Suppl.\ }{\bf #1}}
\newcommand{\AP}[1]{{\it Ann.\ Phys.\ }{\bf #1}}

\end{document}